\documentclass[%
 reprint,
superscriptaddress,
showpacs,preprintnumbers,
 amsmath,amssymb,
 aps,
prl,
]{revtex4-1}

\usepackage{amsmath}    
\usepackage{graphicx}   
\usepackage{verbatim}   
\usepackage{color}      
\usepackage{subfigure}  
\usepackage{hyperref}   
\usepackage{units}      
\usepackage{tabularx}

\newcommand{\ket}[1]{\left| #1 \right \rangle}
\newcommand{\Cr}{${}^{52}\mathrm{Cr}$ }
\newcommand{\Sdrei}{{}^{7}\mathrm{S}_{3} }

\newcommand{\Pdrei}{{}^{7}\mathrm{P}_{3} }

\newcommand{\molL}{\vec{L}}	 
\newcommand{\molS}{\vec{S}}	 
\newcommand{\molSqn}{S}
\newcommand{\molj}{\vec{j}}  

\newcommand{\atLqn}{\tilde{L}}  			
\newcommand{\atSqn}{\tilde{S}}

\newcommand{\atmj}{m_{\tilde{j}}}   

\begin{document}

\title{Photoassociation of spin polarized Chromium}

\author{Jahn R\"uhrig} 
\affiliation{5. Physikalisches Institut and Center for Integrated Quantum Science and Technology, Universit\"at Stuttgart, Pfaffenwaldring 57 D-70569 Stuttgart, Germany}
\author{Tobias B\"auerle} 
\affiliation{5. Physikalisches Institut and Center for Integrated Quantum Science and Technology, Universit\"at Stuttgart, Pfaffenwaldring 57 D-70569 Stuttgart, Germany}
\author{Paul S. Julienne}
\affiliation{Joint Quantum Institute, University of Maryland and NIST, College Park, Maryland 20742, USA}
\author{Eite Tiesinga}
\affiliation{Joint Quantum Institute, National Institute of Standards and Technology and University of
Maryland, 100 Bureau Drive, Mail Stop 8423, Gaithersburg, Maryland 20899, USA}
\author{Tilman Pfau}
\affiliation{5. Physikalisches Institut and Center for Integrated Quantum Science and Technology, Universit\"at Stuttgart, Pfaffenwaldring 57 D-70569 Stuttgart, Germany}

\date{\today}

\pacs{32.80.-t, 32.80.Pj, 34.50.Rk, 31.50.Df, 37.10.De, 34.10.+x, 32.70.Cs}


\date{\today}

\begin{abstract}
We report the homonuclear photoassociation (PA) of ultracold \Cr atoms in an optical dipole trap. This constitutes the first measurement of PA in an element with total electron spin $\atSqn>1$.
Although Cr, with its $\Sdrei$ ground and ${}^{7}\mathrm{P}_{4,3,2}$ excited states, is expected to have a complicated PA spectrum we show that a spin polarized cloud exhibits a remarkably simple PA spectrum when circularly polarized light is applied. 
Over a scan range of \unit[20]{GHz} below the $\Pdrei$ asymptote we observe two distinct vibrational series each following a LeRoy-Bernstein law for a $C_3 / R^{3}$ potential with excellent agreement. 
We determine the $C_3$ coefficients of the Hund's case c) relativistic adiabatic potentials to be \unit[-1.83$\pm$0.02]{a.u.} and \unit[-1.46$\pm$0.01]{a.u.}. 
Theoretical non-rotating Movre-Pichler calculations enable a first assignment of the series to $\Omega=6_u$ and $5_g$ potential energy curves. 
In a different set of experiments we disturb the selection rules by a transverse magnetic field which leads to additional PA series.
\end{abstract}

\maketitle
Proposed in 1987 by Thorsheim et al. \cite{PhysRevLett.58.2420} \textit{photoassociation} (PA) has become a valuable source of precise spectroscopic information as well as a tool to create electronically excited molecules from two, initially free, atoms. 
PA has been used to determine scattering lengths \cite{PhysRevLett.74.1315,PhysRevLett.74.3764,Tiesinga1996,PhysRevLett.93.123202} and accurate values for atomic radiative lifetimes \cite{0295-5075-35-2-085,PhysRevLett.93.123202,PhysRevA.51.R871,PhysRevA.55.R1569}.
The underlying principle is that two colliding atoms can absorb a photon to create a molecule in a well defined quantum state when the photon-detuning matches the energy of an electronically excited bound state. 
Because the principle is independent of the specific choice of the atoms PA can be used to form  homonuclear as well as heteronuclear dimers. 
In two species gases heteronuclear PA has been used to create molecules with large electric dipole moments. An overview on heteronuclear PA can be found in \cite{doi:10.1021/cr300215h}.
For homonuclear PA resonant dipole-dipole interaction leads to a long ranged $C_3 / R^{3}$ interaction potential. 
Homonuclear PA has been shown in alkaline metal \cite{PhysRevA.51.R871,0295-5075-35-2-085,PhysRevA.55.R1569,PhysRevLett.74.3764,PhysRevLett.80.4402}, alkaline earth metal \cite{PhysRevLett.85.2292,SrPA,PhysRevLett.96.203201}, metastable noble gas \cite{Herschbach2000, PhysRevA.83.052516, PhysRevA.81.013407}  systems as well as in Yb \cite{PhysRevLett.93.123202}. 
As a result of small atomic electronic orbital angular momenta $\atLqn$ and small atomic spins $\atSqn$ all the aforementioned elements have in common that their number of ground and excited states is assessable (e.g. for ${}^{2}\mathrm{S} + {}^{2}\mathrm{P}$ in Na there are 16 adiabatic potentials). 
This contrasts to chromium where $\atSqn=3$ leads to a high multiplicity and thus to a large number of possible ground and excited states. The $\Sdrei + {}^{7}\mathrm{P}_{4,3,2}$ in Cr already has 166 adiabatic potentials that have to be taken into account. 
From these 166 potentials 56 dissociate to the $\Sdrei + \Pdrei$ asymptote which is why Cr is expected to have a complicated PA spectrum. 

\noindent
Here we utilize demagnetization cooling \cite{fattori06,Volchkov2014,ruehrig2015}, an optical cooling method for dipolar gases, as a spectroscopy tool. 
We find that the combination of an almost spin-polarized cloud and $\sigma^-$ light leads to a stunningly simple and comprehensible PA spectrum. 
This scheme should be applicable to other highly dipolar elements like Er and Dy which have recently gained considerable interest \cite{Lev2011,Aikawa12,Ferlaino2014,PhysRevX.5.041029,kadau15}. 
\noindent
The scheme also enables the optical creation of magnetic diatomic molecules and provides an exciting perspective to create $\mathrm{Cr}_2,\mathrm{Er}_2,\mathrm{Dy}_2$ dimers - molecules with magnetic moments of up to \unit[20]{$\mu_B$}. 
An extension to a two color PA \cite{nikolov2000} or stimulated Raman PA scheme \cite{Wynar2000,beaufils08} to efficiently create highly magnetic cold ground state molecules seems feasible and complements the approach using Feshbach resonances \cite{FerlainoFeshbachMol2015}. 
Chromium with its 6 unpaired electrons, is considered to be a problem that is particularly hard to treat theoretically. 
Calculations for small internuclear separations exhibited large errors when compared to experimental data \cite{sadeghpour_FBR,sadeghpour_C6,Andersson1995}. 
The PA spectrum obtained here also holds valuable information of short range potentials and quantum defects which may help to develop superior theoretical models. 

We started our measurements by loading $\sim 1.5\cdot 10^{6}$ bosonic \Cr atoms with a temperature of \unit[90]{$\mu K$} in a single beam optical dipole trap (ODT) (trapping frequencies: $\omega_x = \omega_y =  2 \pi \cdot \unit[5.5]{kHz}$ and $\omega_z = 2\pi \cdot \unit[40]{Hz}$). 
As a consequence of the loading mechanism \cite{falkenau11,volchkov2013} the atoms were initially spin polarized in the lowest Zeeman substate $\atmj=-3$ of the $\Sdrei$ ground state, which is a dark state for the $\sigma^-$ polarized optical pumping light. 
Demagnetization cooling \cite{ruehrig2015,Volchkov2014} was started by lowering the homogeneous offset magnetic field suddenly to $B_x \approx \unit[300]{mG}$. The transversal magnetic fields $B_y$ and $B_z$ were of negligible size and were optimized separately to maximize the $\sigma^-$ polarization purity. 
Simultaneously we applied the \unit[427]{nm} optical pumping light with a detuning $\Delta=\omega - \omega_A$ and a constant optical pumping scattering rate of $\Gamma_{SC}=2 \pi \cdot \unit[100]{Hz}$ for the cooling time $t_{cool}=\unit[4]{s}$. Here $\omega$ is the laser frequency and $\omega_A$ is the frequency of the atomic $\Sdrei \leftrightarrow \Pdrei$ transition.
The linewidth of the optical pumping laser has been experimentally determined to be well below \unit[30]{kHz}. 
$\Gamma_{SC}$ is chosen such that it exceeds the dipolar relaxation rates for $\Delta \atmj=\pm 1$ i.e. the regime can be regarded to be saturated \cite{Volchkov2014}. 
The dipolar relaxations thermally excite atoms from $\atmj=-3$ to $\atmj=-2$ at the expense of kinetic energy as they couple the internal degree of freedom (spin) to the external degree of freedom (angular momentum). 
Atoms in the $\atmj=-2$ state couple to the optical pumping light and are thus repumped via the $\Pdrei$ excited state. Subsequent thermalization then leads to a net cooling effect of the whole sample. 
Finally we determined the number of atoms $N$ and the temperature $T$ by absorption imaging. 
We repeated the measurement procedure described above scanning the laser detuning between 0 and \unit[-20]{GHz} relative to the atomic resonance. 

When the detuning matches the energy difference to a molecular vibrational level we observe trap loss as well as heating of the cloud.
\begin{figure}
	\centering
		\includegraphics[width=0.48 \textwidth]{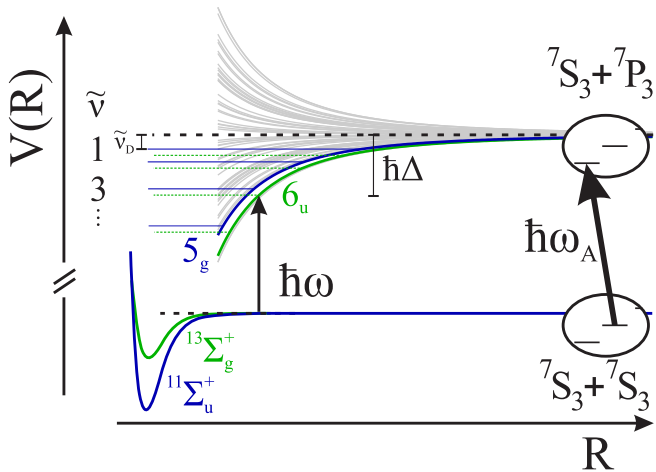}
	\caption{Sketch of the relevant molecular potentials. 
Ground state scattering channels are constrained to ${}^{11}\Sigma_u^+$ and ${}^{13}\Sigma_g^+$. 
For $R\rightarrow\infty$, $\Sdrei + \Sdrei$ and $\Sdrei + \Pdrei$ depict the separated atoms asymptotes. 
The insets show the $\atmj=-3,-2$ Zeeman-levels of the $\Sdrei$ and $\Pdrei$ states. 
At intermediate distances the $C_3$ potentials dominate the transition energies. 
Photoassociation can occur when the laser detuning $\Delta$ matches the energy of a vibrational level $E(\tilde{\nu})$. The main contribution to the PA signal stems from the $6_u$ (green) \& $5_g$ (blue) relativistic adiabatic potentials. }
	\label{fig:Sketch_MolecularLevels_6b}
\end{figure}
In previous publications \cite{fattori06,hensler05,Volchkov2014} we have treated the optical pumping process, which is vital for demagnetization cooling in dipolar gases, in a separated atom picture. This corresponds to the $R \rightarrow \infty$ limit depicted in the insets in Figure \ref{fig:Sketch_MolecularLevels_6b}. 
To understand PA during demagnetization cooling we have to extend this model to a diatomic molecular picture.
The ground and excited state are then represented by a molecular state and the treatment of the problem can be simplified by choosing an appropriate basis \footnote{From a mathematical point of view all bases are equal. 
Nevertheless it is worth to carefully select the appropriate basis to simplify the calculation of the molecular potentials.} - i.e. the appropriate Hund's coupling case. 
For the ground state potential we will use the Hund's case a) basis and thus nonrelativistic Born-Oppenheimer potentials ${}^{2\molSqn+1}\Lambda_{g/u}^{\pm}$, where $\molSqn$ labels the quantum number of the total molecular electronic spin $\molS$, $\Lambda$ is the projection of the total molecular electronic orbital angular momentum $\molL$ on the internuclear axis, $g/u$ labels the inversion symmetry and $\pm$ is the symmetry according to reflection at a plane containing the internuclear axis. 
To account for the strong spin-orbit coupling the excited state is described best by Hund's case c) - i.e. relativistic adiabatic potentials $\Omega_{g/u}^{\pm}$. 
In this coupling case $\Omega$, which is the projection of the total electronic angular momentum $\molj=\molL+\molS$ on the internuclear axis, is a good quantum number.

Atoms in the $\atmj=-3$ are in a dark-state for the $\sigma^-$ polarized optical pumping light and the population of $\atmj>-3$ states is always close to zero due to the presence of the optical pumping light \cite{Volchkov2014}. 
Once dipolar relaxations thermally excite an atom to $\atmj=-2$ this atom is most likely to have an $\atmj=-3$ atom within close proximity.
This suggests the $\ket{\Sdrei,\atmj=-3} +\ket{\Sdrei,-2}$ collisional input channel as the source of the observed PA signal. 
We verify this assumption by the comparison of two-body loss coefficients and timescales of trap loss. 
Theoretical two-body PA rate coefficients are expected to be on the order of $L_2 \approx 5 \cdot \unit[10^{-17}]{m^3/s}$ \cite{PhysRevA.92.022709}.
If we assume both atoms in $\atmj=m_A=m_B=-3$ and two-body losses $\beta_{m_A,m_B} \cdot n_{m_A} n_{m_B}$ with $\beta_{-3,-3}=L_2$ trap loss would occur on the ms timescale. 
As we do not observe such rapid decay we exclude this PA channel. 
In addition to that we observe a perfect agreement of $L_2$ and $\beta_{-3,-2}$ obtained by a numerical loss simulation that accounts for cooling of the cloud. 
From these premisses we conclude that the incoming channel for the ground state is a collision of one atom in $\atmj=-3$ and one atom in $\atmj=-2$. 
There are two molecular Born-Oppenheimer ground state potentials that can have this spin projection of $-5$ which are the ${}^{11}\Sigma_{u}^{+}$ and the ${}^{13}\Sigma_{g}^{+}$ state \cite{Andersson1995}. 
Figure \ref{fig:Sketch_MolecularLevels_6b} depicts these two ground state potentials as solid green and blue lines at the bottom. 
The dependance on the interatomic distance $R$ is given by short ranged Van der Waals interaction which alters the potential energy difference significantly when $R$ is smaller than the Van der Waals length scale $R_{B}$ - i.e. for detunings that were not experimentally accessible to us. 
Therefore, laser detunings are a direct measure for the excited state potential energy curve.  
The selection rules for dipole allowed transitions constrain excited states which are addressable from the ${}^{11}\Sigma_{u}^{+}$ and ${}^{13}\Sigma_{g}^{+}$ states. 
The odd parity of the electric dipole-moment operator requires a $g/u$ parity change so that the ${}^{13}\Sigma_{g}^{+}$ (${}^{11}\Sigma_{u}^{+}$) couples to $\Omega_u$ ($\Omega_g$) 

\begin{figure}
	\includegraphics[trim=0.6cm 2.2cm 1.0cm 2cm, clip, width=0.49 \textwidth]{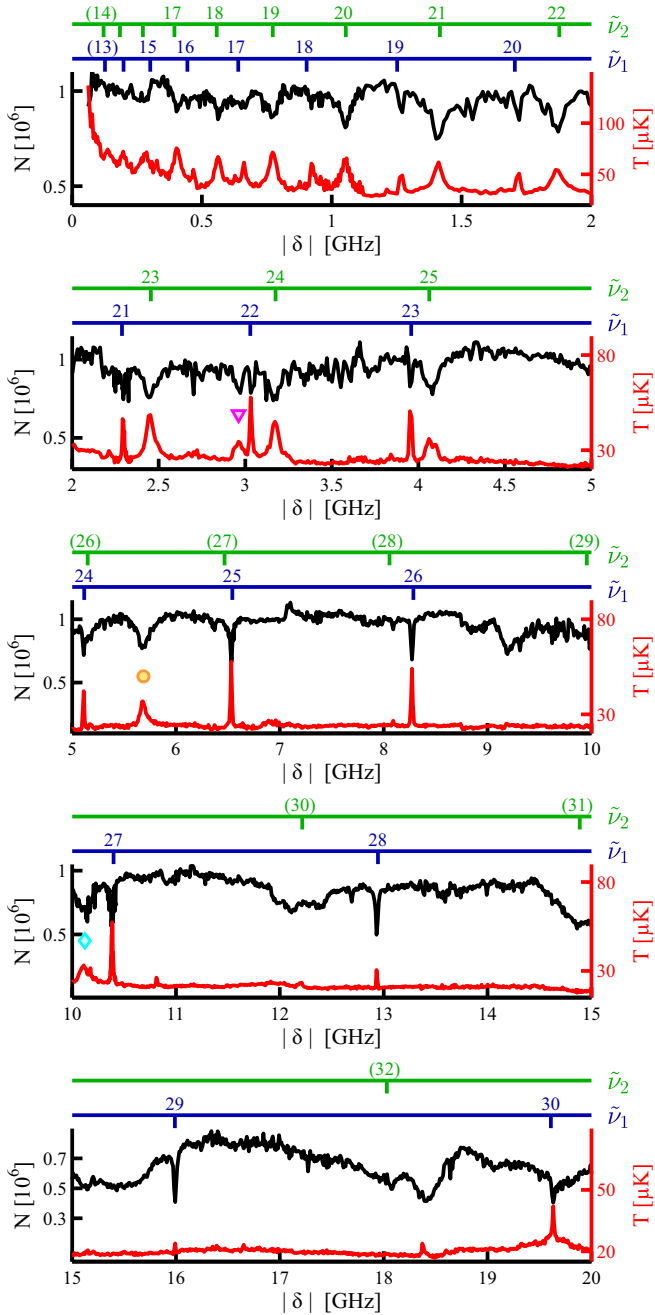}  
	\caption{
	Experimental PA spectrum of \Cr for constant $\Gamma_{sc}$. 
	The black (red) curve shows the observed number of atoms $N$ (cloud temperature $T$) after 4 seconds of demagnetization cooling. 
We find two distinct vibrational series and display their respective vibrational quantum numbers $\tilde{\nu}_{1,2}$ in additional axes on top of each detuning-segment. 
}
	\label{fig:spectrum_gf}
\end{figure}

Figure \ref{fig:spectrum_gf} depicts our PA spectrum as the number of atoms $N$ (top black data) and the temperature $T$ (bottom red data) after \unit[4]{s} of demagnetization cooling as a function of the reduced detuning $\delta=\Delta/ 2\pi$. 
Both quantities are displayed as the visibility for small (large) detunings is more distinct in the temperature $T$ (number of atoms $N$). 
The additional axes on top of each sub-plot show the vibrational quantum number $\tilde{\nu}_i$, where $i=1,2$ is the index of a series of resonances belonging to the same $C_3$ coefficient. 

To obtain the $\tilde{\nu}_i$ we determine the membership of each observed resonance to a series with help of a Fourier transform $\mathcal{F}(T(\delta^{1/6}))$. 
The $1/6$ power leads to equidistant spacing of resonances of the same series and the Fourier transform $\mathcal{F}$ reveals the number of different series. 
After this discernible consecutive resonances of the same series were fitted to a Lorentzian function determining the position $\delta_{exp}$ and width $\gamma(\tilde{\nu}_i)$. 
All fits were done on the temperature data. The consecutive $\delta_{exp}$ were then fitted to a LeRoy-Bernstein equation (LBE) \cite{PhysRevLett.85.2292,Bouloufa.2009}
\begin{equation}
E(\tilde{\nu}_i)=h \cdot \delta=-X_0 \left( \tilde{\nu}_i -1 + \tilde{\nu}_D \right)^{6},
\label{eq:LeRoyBernstein}
\end{equation}
where $\tilde{\nu}_D$ is the non-integer spacing from the last bound state, which has $\tilde{\nu}=1$, to the  threshold as depicted in Figure \ref{fig:Sketch_MolecularLevels_6b}. 
The LBE counts the vibrational states $\tilde{\nu}$ starting from the dissociation limit rather than the lowest lying state. 
The proportionality constant \cite{PhysRevLett.85.2292,Bouloufa.2009}
\begin{equation}
X_0=\left[  \frac{\Gamma(4/3)}{ 2 \sqrt{2 \pi} \Gamma(5/6) } \right]^6 \, \frac{h^6}{\mu^3 C_3^2}
\label{eq:X0}
\end{equation}
relates the resonance positions to the $C_3$ coefficients. In Eq. (\ref{eq:X0}) $\Gamma(z)$ is the Euler Gamma-function and $\mu$ is the reduced mass. 

\begin{figure}[t]
	\centering
		\includegraphics[width=0.45\textwidth]{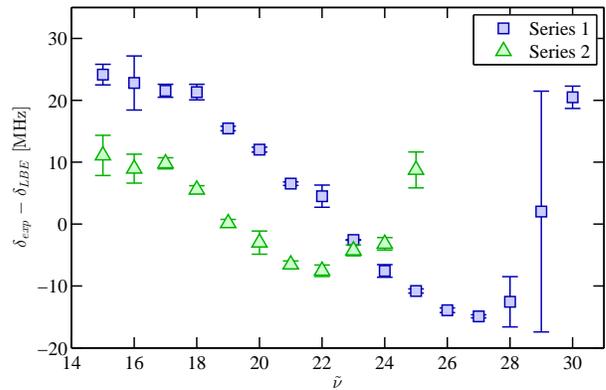}
	\caption{Residuals $\delta_{exp}-\delta_{LBE}$ of the LeRoy-Bernstein fit. The errorbars mark the standard deviations from the Lorentzian fit on the resonances. }
	\label{fig:LBdeviation}
\end{figure}

For $|\delta|<\unit[200]{MHz}$ the resonance spacing is smaller than the resonance width and demagnetization cooling is inefficient and slow which results in a high final temperature. 
The PA resonances can be resolved for $|\delta|\geq \unit[200]{MHz}$. 
A prominent feature of the spectrum are the different widths $\gamma(\tilde{\nu}_i)$ of the different series. 
\begin{figure}
	\centering
		\includegraphics[width=0.45\textwidth]{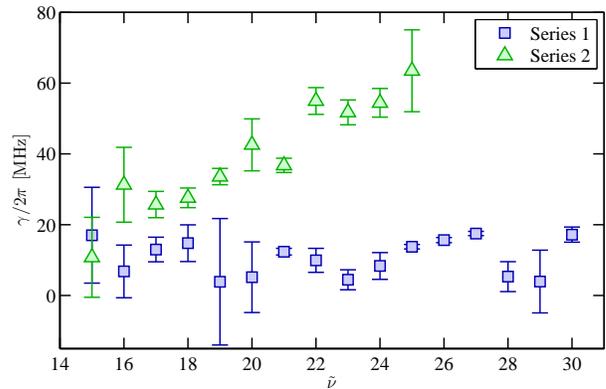}
	\caption{Observed linewidths $\gamma(\tilde{\nu}_i)$ as a function of the vibrational quantum number $\tilde{\nu}_i$. For series 1 (blue data) the linewidth is $\sim \unit[10]{MHz}$. For series 2 (green data) the width increases with $\tilde{\nu}_2$.}
	\label{fig:ResWidths}
\end{figure}
Figure \ref{fig:ResWidths} depicts the widths for series 1 (blue squares) and series 2 (green triangles). 
While $\gamma(\tilde{\nu}_1)$ stays constant at a value of $\sim 2 \gamma_{nat}$, where $\gamma_{nat}$ is the  atomic natural linewidth, $\gamma(\tilde{\nu}_2)$ the width of series 2 gradually increases with $\tilde{\nu}_2$ until the series completely vanishes for $\tilde{\nu}_2>25$.
Even though the series vanishes Fig. \ref{fig:spectrum_gf} continues to show the resonance positions extrapolated from the LBE fit in brackets. For very large detunings $|\delta|> 11 $ GHz series 2 reemerges as very broad atom loss resonances. 
We attribute this broadening to the laser power $P$ that was increased quadratically $P\propto \Gamma_{SC} \Delta^2$ with the detuning to keep the (atomic) scattering rate $\Gamma_{SC}$ constant. 
Within this model also the vanishing of series 2 can be understood in terms of a vanishing Franck-Condon factor due to the first node of the ground state scattering wave function \cite{Burnett96,ruehrig2015}. 
\begin{figure}
	\centering
		\includegraphics[width=0.49\textwidth]{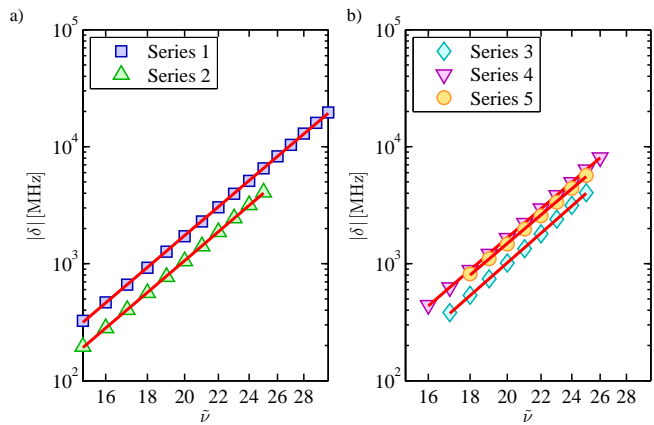}
	\caption{Double logarithmic representation of the observed resonance positions $\left( \delta_{exp}(\tilde{\nu}_i) \right)$ and the vibrational quantum number $\tilde{\nu}_i$ a) for series 1,2 (blue, green data) and b) series 3-5 (cyan, magenta, orange). Only resonances shown here are used in the LBE analysis to obtain the vibrational quantum numbers $\tilde{\nu}_i$. The error bars for series 1,2 are too small to be visible. Linear fits (red lines) retrieve the $\tilde{\nu}^6$ behavior.}
	\label{fig:loglog_allseries}
\end{figure}
Figure \ref{fig:loglog_allseries} a) depicts the resonance positions and vibrational quantum numbers in a double-logarithmic representation. The linear fits (red lines) retrieve the $\tilde{\nu}^6$ behavior and thus verify the LBE modeling. 
Deviations from the pure $R^{-3}$ scaling are of fundamental interest and have been studied by many authors \cite{comparat2004,Bouloufa.2009,PhysRevLett.93.123202,PhysRevA.54.R5,LeRoy1980}. 
In Figure \ref{fig:LBdeviation} we show the residuals $\delta_{exp}-\delta_{LBE}$ of the LBE fit. 
The errorbars mark standard deviations obtained from the Lorentzian fit of the resonances. 
We observe no monotonically increasing deviation from the LBE although there is a similar qualitative shape of the residuals.
\begin{table}[b]
  \centering
  \caption{Comparison of experimental and theoretical $C_3$ coefficients.}
		\begin{tabular}{c| p{2cm} p{0.8cm} p{1.8cm} p{1cm} p{1cm} }
    series & $C_3^{LBE}$ [a.u.] & $\tilde{\nu}_D$ & $C_3^{theo}$ [a.u.] & $\Psi_e$ & $\Psi_g$\\
    \hline
    $\,$  & $\,$  & $\,$  & $\,$  & $\,$  & $\,$\\
    1     & $-1.46\pm0.01$ & 0.91  & $-1.54\pm0.01$ & $5_g$ & ${}^{11}\Sigma_u^+$ \\
    2     & $-1.83\pm0.02$ & 0.83  & $-1.85\pm0.01$ & $6_u$ & ${}^{13}\Sigma_g^+$ \\
    \end{tabular}
  \label{tab:C3}
\end{table}
Table \ref{tab:C3} summarizes the LBE fit results and compares the $C_3$ coefficients to our non-rotating Movre-Pichler \cite{movrepichler1977,RevModPhys.78.483} potential energy curve calculations shown in Fig. \ref{fig:Sketch_MolecularLevels_6b}. We observe good agreement of experimental and theoretical $C_3$ coefficients for the $6_u$ and $5_g$ relativistic adiabatic potentials. 
Resonances which do not belong to series 1 or 2 can observed in Fig. \ref{fig:spectrum_gf} where they are marked with a symbol of their respective series (see Fig. \ref{fig:loglog_allseries}). 
The assignment of these resonances to a specific series will be resolved in the following.

In a different set of experiments we disturb the strict selection rules by applying a transverse magnetic field.
This resulted in the appearance of additional resonances. 
The visibility of these additional resonances is in general not as good as for series 1,2 because they often appear as unresolved side-peaks. 
We extracted the positions by hand and applied the LBE analysis as explained above. 
From this analysis we obtain three additional series of resonances that also include all the resonances that did not belong to series 1 or 2 - e.g. the \unit[5.7]{GHz} resonance. 
Figure \ref{fig:loglog_allseries} b) depicts the double-logarithmic representation of $|\delta|$ and $\tilde{\nu}_i$ for series 3-5. 
As before we observe a perfect agreement with the LBE and extract $C_3$ coefficients \unit[$-1.70\pm0.04$]{a.u.} ,\unit[$-1.47\pm0.01$]{a.u.} and  \unit[$-1.55\pm0.06$]{a.u.}. 
We explain the appearance of additional resonances when tilting the magnetic field with the excitation to $\Pdrei$ states with $\atmj>-3$ which allows the formation of states with smaller $\Omega$. 
It remains however an open question why particular resonances of series 3-5 appear in the case of not-tilted magnetic fields. 

In conclusion, we have observed photoassociation in the highly magnetic atomic species \Cr. In the accessible scan range of 20 GHz below the dissociation limit we observed more than 50 bound states belonging to 5 series of resonances. 
We experimentally determined the $C_3$ coefficients of the 5 series by a LeRoy-Bernstein analysis and compared them to calculations of non-rotating relativistic adiabatic potentials. 
This enabled us to assign two of the series to specific $\Omega$ states. 

We thank A. Griesmaier for his contributions in the earlier stages of the experiment. 
This work was supported by the DFG under Contract No. PF381/11-1.

%

\end{document}